\documentclass{cjpsuppl}
\usepackage{graphicx,epsfig}

\begin{document}

\newcommand{\beqn}{\begin{eqnarray}}
\newcommand{\eeqn}{\end{eqnarray}}
\newcommand{\ra}{\rightarrow}
\newcommand{\np}{Nucl.\,Phys.\,}
\newcommand{\pl}{Phys.\,Lett.\,}
\newcommand{\pr}{Phys.\,Rev.\,}
\newcommand{\prl}{Phys.\,Rev.\,Lett.\,}
\newcommand{\prep}{Phys.\,Rep.\,}
\newcommand{\nuclinst}{{\em Nucl.\ Instrum.\ Meth.\ }}
\newcommand{\annp}{{\em Ann.\ Phys.\ }}
\newcommand{\intjmp}{{\em Int.\ J.\ of Mod.\  Phys.\ }}

\def\mchi{m_{\chi^+}}
\def\neuto{\tilde{\chi}_1^0}
\def\mneuto{m_{\tilde{\chi}_1^0}}
\def\neutt{\tilde{\chi}_2^0}
\def\neutth{\tilde{\chi}_3^0}
\def\ma{M_A}
\def\mstau{m_{\tilde\tau}}
\def\msne{m_{\tilde\nu}}
\def\msnee{m_{{\tilde\nu}_e}}
\def\mh{M_h}
\def\noi{\noindent}
\def\sinb{\sin\beta}
\def\cosb{\cos\beta}
\def\sinbb{\sin (2\beta)}
\def\cosbb{\cos (2 \beta)}
\def\tgb{\tan \beta}
\def\tgbt{$\tan \beta\;\;$}
\def\tgbsq{\tan^2 \beta}
\def\sinal{\sin\alpha}
\def\cosal{\cos\alpha}
\def\stop{\tilde{t}}
\def\sto{\tilde{t}_1}
\def\stt{\tilde{t}_2}
\def\stl{\tilde{t}_L}
\def\str{\tilde{t}_R}
\def\msto{m_{\sto}}
\def\mstosq{m_{\sto}^2}
\def\mstt{m_{\stt}}
\def\msttsq{m_{\stt}^2}
\def\mt{m_t}
\def\mtsq{m_t^2}
\def\sint{\sin\theta_{\stop}}
\def\sintt{\sin 2\theta_{\stop}}
\def\cost{\cos\theta_{\stop}}
\def\sintsq{\sin^2\theta_{\stop}}
\def\costsq{\cos^2\theta_{\stop}}
\def\mqtt{\M_{\tilde{Q}_3}^2}
\def\mutt{\M_{\tilde{U}_{3R}}^2}
\def\sbottom{\tilde{b}}
\def\sbo{\tilde{b}_1}
\def\sbt{\tilde{b}_2}
\def\sbl{\tilde{b}_L}
\def\sbr{\tilde{b}_R}
\def\msbo{m_{\sbo}}
\def\msbosq{m_{\sbo}^2}
\def\msbt{m_{\sbt}}
\def\msbtsq{m_{\sbt}^2}
\def\mt{m_t}
\def\mtsq{m_t^2}
\def\selectron{\tilde{e}}
\def\seo{\tilde{e}_1}
\def\set{\tilde{e}_2}
\def\sel{\tilde{e}_L}
\def\se1{\tilde{e}_1}
\def\ser{\tilde{e}_R}
\def\mseo{m_{\seo}}
\def\mseosq{m_{\seo}^2}
\def\mset{m_{\set}}
\def\msetsq{m_{\set}^2}
\def\msel{m_{\sel}}
\def\mser{m_{\ser}}
\def\mse1{m_{\se1}}
\def\me{m_e}
\def\mesq{m_e^2}
\def\snu{\tilde{\nu}}
\def\snue{\tilde{\nu_e}}
\def\set{\tilde{e}_2}
\def\snul{\tilde{\nu}_L}
\def\msnue{m_{\snue}}
\def\msnuesq{m_{\snue}^2}
\def\smuon{\tilde{\mu}}
\def\smur{\tilde{\mu}_R}
\def\msmul{m_{\smul}}
\def\msmulsq{m_{\smul}^2}
\def\msmur{m_{\smur}}
\def\msmursq{m_{\smur}^2}
\def\stau{\tilde{\tau}}
\def\stauo{\tilde{\tau}_1}
\def\staur{\tilde{\tau}_R}
\def\mstauo{m_{\stauo}}
\def\mstauosq{m_{\stauo}^2}
\def\mstaut{m_{\staut}}
\def\mstautsq{m_{\staut}^2}
\def\mtau{m_\tau}
\def\mtausq{m_\tau^2}
\def\gluino{\tilde{g}}
\def\mgluino{m_{\tilde{g}}}
\def\mchi{m_\chi^+}
\def\neuto{\tilde{\chi}_1^0}
\def\mneuto{m_{\tilde{\chi}_1^0}}
\def\neutt{\tilde{\chi}_2^0}
\def\mneutt{m_{\tilde{\chi}_2^0}}
\def\chargop{\tilde{\chi}_1^+}
\def\chargopm{\tilde{\chi}_1^\pm}
\def\mchargo{m_{\tilde{\chi}_1^+}}
\def\chargom{\tilde{\chi}_1^-}
\def\bino{\tilde{b}}
\def\wino{\tilde{w}}
\def\m0{M_0}
\def\mhf{M_{1/2}}
\def\r12{r_{12}}
\def\r32{r_{32}}
\def\bsgamma{b\ra s\gamma}
\def\bsmu{B_s\ra \mu^+\mu^-}
\def\gmuon{$(g-2)_\mu$}
\def\feynhiggs{{\tt FeynHiggs}}
\def\micro{{\tt micrOMEGAs}}
\def\mhf{M_{1/2}}
\def\suspect{{\tt Suspect}}
\def\softsusy{{\tt SOFTSUSY}}
\def\xenon{{\tt Xenon}}
\def\zeplin{{\tt Zeplin}}
\def\zepliniv{{\tt ZeplinIV}}
\def\edelweiss{{\tt Edelweiss}}
\def\genius{{\tt Genius}}
\def\cdms{{\tt CDMS}}
\def\picasso{{\tt PICASSO}}
\def\simple{{\tt Simple}}
\def\naiad{{\tt Naiad}}
\def\dama{{\tt DAMA}}
\def\epem{e^+e^-}
\def\mbmb{m_b(m_b)}
\def\amu{a_\mu}

\title{Relic density of dark matter in mSUGRA and non-universal SUGRA}
\authori{G.~B\'elanger, F.~Boudjema, A.~Cottrant}
\addressi{LAPTH, 9 Chemin de Bellevue, 74940 Annecy-le-Vieux, France}
\authorii{A.~Pukhov}    \addressii{SINP, Moscow State University, Moscow, Russia}
\authoriii{A.~Semenov}   \addressiii{JINR, Dubna, Moscow Region, Russia}
\authoriv{}    \addressiv{}
\authorv{}     \addressv{}
\authorvi{}    \addressvi{}
\headtitle{Relic density of dark matter in mSUGRA \ldots}
\headauthor{G. B\'elanger} \lastevenhead{G. B\'elanger: Relic
density of dark matter in mSUGRA \ldots} 
\refnum{}
\daterec{6 October 2004;
} \suppl{A}  \year{2004} \setcounter{page}{1}
\maketitle

\begin{abstract}
The  measurements of WMAP  on the relic density of dark matter strongly constrain supersymmetric models. In mSUGRA where the neutralino LSP is mostly a bino only rather fine-tuned models survive.  On the other hand
the relic density upper limit can be easily satisfied in 
models with a Higgsino or wino LSP. 
\end{abstract}

\section{Introduction}

Despite the lack of direct  experimental evidence for
supersymmetry, the minimal supersymmetric standard model (MSSM)
remains one of the most attractive extensions of the standard
model. One of  the nice features of the MSSM with R-parity conservation is that it provides a
natural cold dark matter candidate, the lightest supersymmetric
particle (LSP). Recently, the WMAP satellite has measured
precisely the relic density of cold dark matter, $\Omega
h^2=.1126^{+.0161}_{-.0181}$ (at 2$\sigma$) \cite{wmap}. This 
measurement severely constrains the parameter space of
supersymmetric models. 
This is particularly true in mSUGRA models where  most of the 
scenarios compatible with WMAP require a careful adjustment of parameters \cite{sugra:wmap,Belanger:sugra}.  
Indeed over most of the  mSUGRA parameter space, the LSP is almost purely  bino. As such,  the LSP  annihilates into fermions through the t-channel exchange of a right-handed sfermion.
 This  process is not efficient enough to satisfy the  tight WMAP upper limit on the relic density of dark matter.  Then  one
must appeal to specific mechanisms such as rapid annihilation of
neutralinos via s-channel Higgs exchange 
or coannihilation of neutralinos with other  sfermions
\cite{coan_funnel} to bring down the relic density in the desired range. 
Another possibility is that the LSP has a significant  Higgsino component, then the couplings  of neutralinos to the Higgs or Z is enhanced   and annihilation into gauge bosons or fermion pairs  becomes very efficient. 
 One of  these conditions is satisfied only for narrow strips in the parameter space of the mSUGRA model \cite{sugra:wmap,Belanger:sugra}. One basically finds three allowed region: the coannihilation region which occurs usually when  $\mstauo \approx \mneuto$; the Higgs
funnel region where annihilation via a  heavy pseudoscalar Higgs into fermion pairs dominate  at large $\tan\beta$; the  focus point
 region where at large values of $\m0$ one finds a neutralino with some  Higgsino component.  Note however that
the location of both the focus point region and the Higgs funnel depend strongly on the standard model input parameters, in particular the top quark mass \cite{uncertainty,Allanach:codes}.

The careful tuning of parameters necessary to comply with the upper limit on the relic density from  WMAP within mSUGRA is however not generic of all MSSM models. Models with either a Higgsino or wino LSP generally satisfy easily the WMAP upper limit because of the large cross section for the annihilation of neutralinos  into gauge bosons or fermion pairs \cite{Birkedal:wino}.
Such scenarios can be  found for example in 
SUGRA models with non-universal gaugino masses \cite{Belanger:sugra,nonuni}. 
 Furthermore, in non-universal models,  the relation between the heavy Higgs
mass and the neutralino mass deviates from the mSUGRA prediction and one finds that annihilation through a heavy Higgs exchange can take place even at low
$\tan\beta$. 

 In order to assess the potential of future
 colliders to discover supersymmetry it is therefore important to consider the different categories of models and examine the impact of  the relic density
 constraint.  Here we review the constraints on mSUGRA models showing also the impact of the top quark mass before discussing non-universal SUGRA models.
 Other constraints including direct searches
 as well as  $\bsgamma$ are also briefly mentionned.

\section{Relic density of dark matter}

The computation of the relic density of dark matter amounts to solving the evolution equation for the density of the LSP which depends on the effective annihilation cross section \cite{Belanger:micro13}. Coannihilation processes involving SUSY particles slightly heavier than the LSP can contribute significantly to the effective  cross-section. However, these contributions
are  suppressed by a Boltzmann factor $\propto exp^{-\Delta M/T_f}$
where $\Delta M$ is the mass difference between the NLSP and the
LSP and $T_f$ the decoupling temperature. 
  Here we will present results computed  with the code \micro \cite{Belanger:micro13}. This code
includes all annihilation and coannihilation channels and  takes into account 
 loop corrections in masses as well as in
vertices. Loop corrections  for the Higgs widths (including the 
$\Delta m_b$ correction)  are  also
included.

Within the context of a model defined at the GUT scale such as mSUGRA or non-universal SUGRA,  one
crucial step is the evaluation of the physical spectrum
corresponding to a given model. This is based on the
renormalization group equations  that specify the evolution
of parameters from the high scale to the low scale as well as on
the inclusion of radiative corrections to the physical masses of
sparticles. The theoretical uncertainties in the
prediction of the spectrum can seriously affect the prediction
of observables at low energies and in particular of the relic
density \cite{uncertainty}. 
Here we will mainly present results obtained with the spectrum calculator \softsusy1.8.6
\cite{Allanach:softsusy} interfaced with \micro~1.3 through  the SUSY Les Houches Accord \cite{SLHA}.

\section{Constraints on mSUGRA models}

\begin{figure}[tbhp]
\begin{center}
\vspace{-1.2cm}
\includegraphics[width=14cm,height=8.cm]{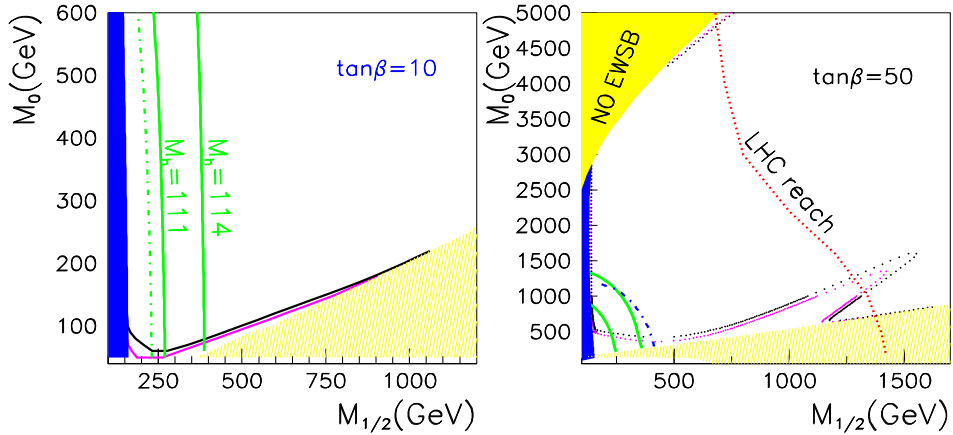}
\vspace{-1.9cm} \caption{\label{sugra_mt175}{\em Allowed regions
in the $\m0-\mhf$ plane  for a) $\tan\beta=10$, $A_0=0$, 
$\mu>0$, $\mt=175$~GeV.
Contours for for $\Omega h^2= .129 (.094)$ (black/pink dots), $\mh=111,114$~GeV (full green), $\mh=111$~GeV,$\mt=179$~GeV (dash-dot).  b) Allowed regions for $\tan\beta=50$, $A_0=0$,  $\mu>0$, $\mt=175$~GeV. The contour for $\bsgamma=2.25\times 10^{-4}$ is also shown (dash-dot). The LHC reach for ${\cal L}=100fb^{-1}$ is reproduced from \cite{Baer:reach} 
}} \vspace{-0.5cm}
\end{center}
\end{figure}

We first remark that within mSUGRA the so-called
bulk region (at low $\mhf$ and $\m0$) that used to be one of the
favoured regions  has considerably shrunk
after the much tighter upper limit on the relic density from WMAP.
This is because in  the bulk region the LSP is almost a pure bino.
 The annihilation  of binos in fermion pairs 
is not efficient enough to satisfy the  upper limit on
the relic density of dark matter.  One needs some additional
contribution from the coannihilation or Higgs exchange channels. The important contribution of the latter at large
$\tan\beta$   explains why
the  region allowed at  low $\m0-\mhf$ is more important, see Fig.~\ref{sugra_mt175}. 
Note that the low $\m0-\mhf$ region is also
constrained from the Higgs mass limit at least for intermediate $\tan\beta$ and by the measurement of $\bsgamma$ when  $\tan\beta>35$.
We are then left mainly with  three regions: coannihilation, focus point and
Higgs funnel. Results for the case $\mt=175$~GeV are presented in each of these regions.

In mSUGRA, the coannihilation region designates the region at
 low $\m0$, where the lightest $\stauo$ is the NLSP. 
 In addition to neutralino annihilation into fermion pairs, the main coannihilation 
channels are $\neuto\stauo\ra \tau \gamma$ and $\stauo\stauo\ra
\tau\bar\tau$. The relic density
is very sensitive to the NLSP-LSP mass difference. The proper mass
degeneracy for a relic density consistent with WMAP corresponds to
a very narrow strip in the $\m0-\mhf$ plane. In Fig.~\ref{sugra_mt175}, the
allowed region correspond to the  area between the
 $\Omega h^2=.094$ and $\Omega h^2=.129$ contours.    Typically
a mass
difference $\Delta M_{\stau\neuto}\approx 10$~GeV for
$\mneuto\approx 100$~GeV is required while near degeneracy
is necessary for heavier neutralinos, $\mneuto=400$~GeV.
 The mass difference between the LSP and the sleptons of the first
two generations  can be larger.  Eventually, as the
LSP mass increases the coannihilation cross sections also
become too small and one obtains an upper limit on the LSP mass.
 We find for $\tan\beta=10$ that  $\mneuto,\mstauo<448$~GeV,$m_{\tilde q}<2$~TeV
 and $m_{\tilde g}<3$~TeV. All this region should be within reach of the LHC
 \cite{Baer:reach}.
Within mSUGRA, one can also find models where coannihilation  with the $\tilde{t}$-quark takes place\cite{stop}. This occurs only for a large value of $A_0$. Although squarks are rather light, the main decay channel is into $\tilde{t}\ra c\neuto$ making it difficult for searches at LHC.

The Higgs annihilation region includes both a light Higgs
annihilation at low $\mhf$ and a heavy Higgs annihilation  at
large $\tan\beta$. The former occurs only for the LSP mass very
close to  $M_h/2$ since the light Higgs has a very narrow width.
The latter appears  at moderate $\m0-\mhf$ values and 
benefits from the enhanced couplings of the heavy pseudoscalar Higgs  to $b$-quarks.
 The main annihilation   channels are  $\neuto\neuto \ra
b\bar{b},\tau\bar{\tau}$ corresponding to the preferred decay
channels of the heavy Higgs. 
 Due to the heavy Higgs annihilation channel,
neutralinos as heavy as $650$~GeV can give reasonable values for
the relic density. The only condition is that $\ma-2\mneuto < 2 \Gamma_A$.
 This type of model  could be very difficult to
hunt at the LHC as it features a rather heavy  spectrum. Indeed at
the tip of the Higgs funnel, one finds all
 squarks and the gluino in the $2.5-3.2$~TeV range.  Note that  the position of the  Higgs funnel
 is sensitive to the standard model input value used in the spectrum calculator codes, in
particular $m_b(m_b)$ \cite{Gomez:mb} as well as, to a lesser
extent,  the top quark mass \cite{Allanach:houchesrge,uncertainty}. 
In evaluating the relic density in the Higgs funnel, the crucial parameters are $\ma$ and $\mneuto$ \cite{uncertainty}. Then the  LHC can help constrain the relic density in this scenario with an accurate measurement of $\ma$ if  the $A \ra \mu^+\mu^-$ channel can be used \cite{Atlas-tdr}.

\begin{figure}[tbhp]
\begin{center}
\vspace{-1.2cm}
\includegraphics[width=14cm,height=8.cm]{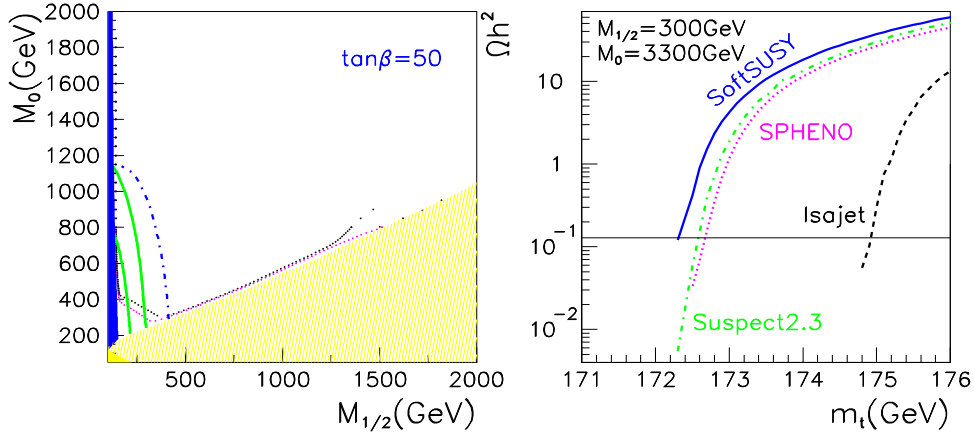}
\vspace{-1.9cm} \caption{\label{fig:mt}{\em a) Allowed regions
in the $\m0-\mhf$ plane  for $\tan\beta=50$, $A_0=0$,
$\mu>0$ and $\mt=179$~GeV. Same labels as Fig.1. b) $\Omega h^2$ vs $\mt$
 for $\tan\beta=10,A_0=0,\mu>0$ and for SoftSUSY1.8.6, Suspect2.3, Spheno2.2.0 and Isajet7.69 \cite{codes}.
}} \vspace{-.5cm}
\end{center}
\end{figure}

The last allowed region is the focus point region, this region is
found at high values of $\m0$ where the value of $\mu$ drops
rapidly. This occurs near the electroweak symmetry breaking border.
When $\mu \approx M_1,M_2$,  the LSP has a significant Higgsino
fraction.
The main annihilation
channels are $\neuto\neuto\ra W^+W^-,ZZ, t\bar{t}$.
Coannihilation channels with heavier neutralinos and/or charginos are also important.
The position of the focus point region is very sensitive to the
value of the top quark mass that enters the RGE and also differs
for different RGE codes,
\cite{uncertainty,Allanach:codes,Allanach:houchesrge} see Fig.~ \ref{fig:mt}b. For
the case of SoftSUSY used here, rather low values of the top quark
mass are necessary to reach the region where $\mu$ drops rapidly when
$\tan\beta=10$.  For $m_t=175$~GeV, one finds a
cosmologically allowed region at large $\m0$ only  for large
$\tan\beta$ as displayed in
Fig.~\ref{sugra_mt175}a. Values
consistent with WMAP for $\Omega h^2$ require a large $\neuto\neuto Z$ coupling, that is not so large a value
for $\mu$. This implies that the chargino and lightest neutralinos
should be rather light. For example,  the chargino is confined to $\mchargo< 375$~GeV for $\tan\beta=50$.  Although the
 chargino and the LSP can be nearly degenerate  this
 occurs mainly when the relic density falls below the WMAP range,
  for $\Omega h^2\approx .1$ typical mass differences are rather around
  $50$~GeV.
 Note that models in the focus point
region feature a very heavy sfermion sector. The LHC will have
little opportunity to discover the squarks, only the gluino could
be accessible. At the same time the rather light gaugino sector
could be probed at a linear collider, extending the reach of the
LHC \cite{Baer:reach}.

\subsection{Theoretical uncertainties:  standard model parameters}

The values of the standard model input parameters 
 play an important role in determining the SUSY spectrum in mSUGRA \cite{uncertainty}.  Therefore they have an impact on the relic density constraint. This is true  especially in  the focus point and the heavy Higgs funnel regions. For one
the heavy Higgs masses, which often provide the only efficient annihilation mechanism also  depend on the top Yukawa as well as on 
$\mbmb$ and $\alpha_s$ \cite{uncertainty,Gomez:mb}. More importantly the  solution of the renormalization group 
equation for the $\mu$ parameter, which drives the Higgsino nature of the LSP,  is extremely sensitive to the top Yukawa \cite{Allanach:codes}.

A heavier top quark, say $m_t=179$~GeV,  means a shift of the focus point region towards higher values of  $\m0$  in fact beyond $\m0=5$~TeV even for $\tan\beta=50$.  For large values of  $\m0$, it is  difficult
 to obtain a converging solution, although the authors of  Ref.~\cite{Baer:2004qq} find the focus point region all
the way up to $M_0=10$~TeV.
 The Higgs funnel moves closer
to the coannihilation region when $\mt=179$~GeV. Indeed the heavy Higgs masses are shifted upwards with the increase of the top Yukawa making it
increasingly difficult to have $\mneuto\approx M_A/2$.  Note that increasing the bottom quark mass has the opposite effect. Finally 
the relaxed bound from the Higgs mass means that one recovers part of 
 the region where light neutralinos/charginos and sleptons can be found.

\section{Beyond mSUGRA}

The nature of the LSP strongly influences the value of the relic
density of dark matter so constraints on SUSY models can differ markedly from the mSUGRA case in models with a Higgsino or wino  LSP. This can be achieved 
by relaxing the gaugino universality condition. 
 For one, the wino
content of the LSP can be significantly increased in models where $M_1/M_2>1$ at the GUT scale. Second the  value of
$\mu$ can be much smaller than predicted in mSUGRA thus increasing the Higgsino fraction of the LSP. This occurs for example 
for models with $M_3/M_2<1$ at the GUT scale. Below we give explicit examples of these two types of models.
Basically one finds the same mechanisms as in mSUGRA for
getting sufficient annihilation/coannihilation of neutralinos but
the
 allowed region in the parameter space of the MSSM shifts significantly
 and no longer requires a fine tuning of parameters especially if one is willing to allow models which are below the WMAP range.
 The direct constraints are of course also modified. 

\vspace{-.1cm}
\subsection{$M_1>M_2$}

\begin{figure}{
 \unitlength=1.1in
\vspace{-.8cm}
\begin{picture}(0.1,2.6)
\put(-.1,0){\epsfig{file=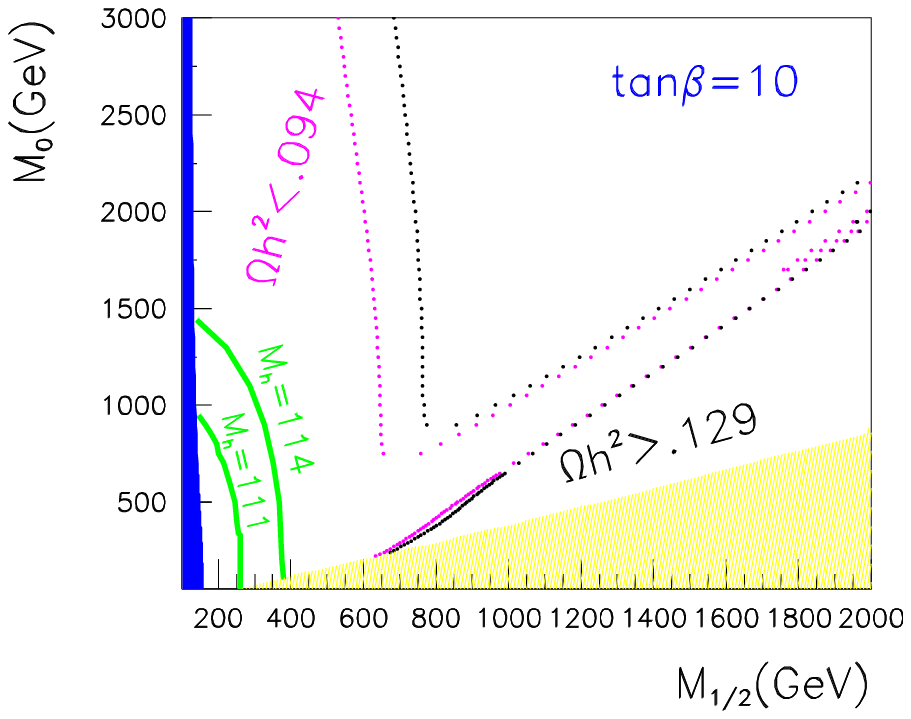,width=8.1cm}}
\end{picture}
\begin{picture}(4.2,2.6)
\put(2.1,0){\epsfig{file=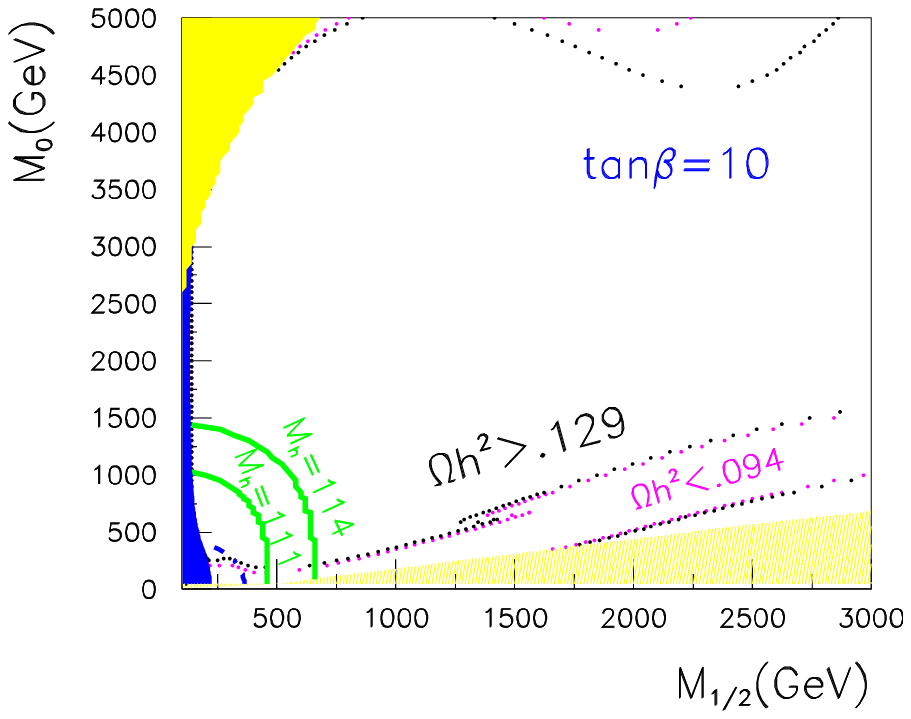,width=8.1cm}}
\end{picture}
\vspace{-1.4cm}
\caption{\it \small  Allowed
regions in the $\m0-\mhf$ plane  for $\tan\beta=10$, $A_0=0$,
$\mu>0$ and $\mt=175$~GeV a) $M_1=1.8M_2$ b) $M_3=M_2/2$. Contours for $\mh=111,114$~GeV (green) and for $\Omega h^2= .129 (.094)$ (black/pink dots). }
\vspace{-0.8cm}
\label{fig:nonuni}}
\end{figure}

Imposing  $M_1> M_2$  at the GUT scale lead to  $M_1\geq M_2$
at the weak scale and a  LSP with  a higher wino content. Then
annihilation of neutralinos into W pairs becomes dominant. Since this
process is much more efficient than the annihilation into fermion
pairs  the relic density often falls below the WMAP range \cite{Birkedal:wino}. 
Consider for example a model with $M_1=1.8M_2$.
This value  corresponds to $M_1\approx M_2$ at the electroweak scale. Then the LSP is a mixed bino/wino state. 
The allowed region in the $\m0-\mhf$ plane differs significantly from the mSUGRA case \cite{Belanger:sugra}. The annihilation
into gauge bosons  occurs and has a large enough rate even when 
the fermion annihilation channels is small because the sfermions are very heavy.  For example in Fig.~\ref{fig:nonuni}a,  the whole region to the
left of the $\Omega h^2=.094$ contour has a relic density below
the WMAP range. Agreement with WMAP then implies a rather heavy
LSP, for example $\mneuto\approx 600$~GeV for $\tan\beta=10$.
The tail in Fig.~\ref{fig:nonuni}a corresponds to  a region where
coannihilation with the chargino becomes significant, this coannihilation proceeds near  the heavy Higgs pole. 
The constraints on this type of models  from the Higgs mass as
well as from precision measurements are not significantly
different than in the universal model. 

As concerns  collider searches,  the spectrum of coloured sparticles  is not dramatically different then in mSUGRA.
However since the WMAP allowed regions are completely different
and in particular include a Higgs funnel at small $\tan\beta$,
models with very heavy   squarks are perfectly acceptable. For
example, for $\m0=\mhf=2$~TeV and $\tan\beta=10$, all squarks are
in the 3-4~TeV range. These are beyond the reach of LHC.
Furthermore, the LSP tends to be much  heavier than in mSUGRA,
more importantly, the small  mass difference between the
chargino and the neutralino  can effect searches.
Had we increased further the ratio $M_1/M_2$ we would have found a
LSP with an even larger wino component. This would increase the
values for the LSP masses compatible with WMAP. For example if $M_1/M_2=2.5$ the LSP as well as the whole supersymmetric spectrum lies above $1.6$~TeV unless there is an additional dark matter component.

\normalsize

\subsection {$M_3 < M_2$, the case $M_3=M_2/2$}

Decreasing the value of $M_3$ while keeping $M_2=M_1$ at the
GUT scale also  means that the relic density constraint  can be
satisfied more easily.
 First, one finds smaller values for $\mu$ which means a LSP with a
significant Higgsino component and more efficient annihilation
channels without requiring  high $\tan\beta$ \cite{nonuni,Belanger:sugra}. 
Second,  one finds significant areas of
parameter space where neutralino annihilation proceeds through 
s-channel Higgs exchange even for moderate values of $\tan\beta$. 
For example in the region of a $\tilde\tau$ NLSP, both stau-coannihilation
 and annihilation into fermion pairs via Higgs exchange contribute to the effective annihilation cross section. As a result, when the $\stau-\neuto$ are nearly degenerate the relic  density
 is already much below the WMAP range unless the LSP is very heavy.
In this model, we also find rapid annnihilation trough a Higgs resonance but this time for the coannihilation
processes $\neuto\neutt,\neuto\neutth\ra
f\bar{f}$  or $\neuto\chargop\ra t\bar{b}$ via charged Higgs exchange. The onset of the charged
Higgs exchange contribution is easily visible as a kink in the
contours of constant relic density in Fig. \ref{fig:nonuni}b. 
For large $\tan\beta$ the relic
density constraint is satisfied in an even larger  portion of 
the parameter space because of the more efficient annihilation/coannihilation through heavy Higgses as well as a  large Higgsino-LSP region. 
Note that in these models  both the Higgs mass and the $\bsgamma$ constraints
become more severe as the squark mass are lower than in the
universal case for a fixed value of $\mhf=M_2|_{GUT}$. 

 A significant fraction  of the
models  with $M_3<M_2$  predict squarks within the range accessible
by the LHC. However the region compatible with WMAP is not completely
accessible even for intermediate values of $\tan\beta$. Because of the
annihilation/coannihilation through the heavy Higgses,  one can find reasonable values for the
relic density  even 
for very heavy neutralinos, hence  a very heavy
supersymmetric spectrum. For example for $\m0= 4$~TeV, $\mhf=3$~TeV,
the spectrum consists of heavy squarks, $m_{\tilde q}=3-4$~TeV,
neutralinos and charginos, $\mneuto\approx\mchargo\approx 820$~GeV
and also heavy Higgses $ m_{H/A}\approx
1.5$~TeV, a difficult task for discovering supersymmetry at
colliders. The only new particle that could be reached at a
collider would be the light Higgs.

\section{Conclusion}

The latest measurements of WMAP on the relic density of dark matter put strong  constraints on supersymmetric models. In particular in mSUGRA 
 very specific relation among sparticle masses and parameters should be satisfied:
 neutralino nearly degenerate with a sfermion,
 neutralino mass near half the mass of a Higgs or 
 neutralino with a significant Higgsino component.
In mSUGRA most of the parameter space would be covered at LHC except for the tip of the Higgs funnel region and  part of the Higgsino LSP region.
In non-universal SUGRA  models, compatibility with the WMAP measurements  imposes similar conditions on the SUSY spectrum than in
mSUGRA. The main new possibility is for a wino LSP. 
However the relaxed constraints  mean that a  significant part of the allowed parameter space is beyond the reach of the LHC.

\providecommand{\href}[2]{#2}\begingroup\raggedright


\begin{thebibliography}{10}

\bibitem{wmap}
D.~N. Spergel {\em et.~al.}, {\em Astrophys. J. Suppl.} {\bf 148} (2003) 175;
C.~L. Bennett {\em et.~al.}, {\em Astrophys. J. Suppl.} {\bf 148} (2003) 1.

\bibitem{sugra:wmap}
H.~Baer {\em et.~al.}, {\em JHEP} {\bf 07} (2002) 050;
J.~R. Ellis {\em et.~al.}, {\em Phys. Lett.} {\bf
  B565} (2003) 176;
U.~Chattopadhyay, A.~Corsetti, and P.~Nath, {\em Phys. Rev.} {\bf D68} (2003)
  035005;
A.~B. Lahanas and D.~V. Nanopoulos, {\em Phys. Lett.} {\bf B568} (2003) 55;


\bibitem{Belanger:sugra}
G.~B\'elanger, F.~Boudjema, A.~Cottrant, A.~Pukhov,  A.~Semenov, hep-ph/0407218.

\bibitem{coan_funnel}
K.~Griest and D.~Seckel, {\em Phys. Rev.} {\bf D43} (1991) 3191;
  M.~Drees and M.~Nojiri, Phys.~Rev. {\bf D47} (1993) 376;
  R.~Arnowitt and P.~Nath, Phys.~Lett.~ {\bf B299}(1993) 58.


\bibitem{uncertainty}
B.~Allanach, {\em et al.}, hep-ph/0410091; B.~Allanach, {\em et al.}, hep-ph/0410049.

\bibitem{Allanach:codes}
B.~C. Allanach, S.~Kraml, and W.~Porod, {\em JHEP} {\bf 03} (2003) 016.

\bibitem{Birkedal:wino}
A.~Birkedal-Hansen and B.~D. Nelson, {\em Phys. Rev.} {\bf D64} (2001) 015008.


\bibitem{nonuni}
H.~Baer {\em et.~al.}, {\em JHEP} {\bf 05} (2002) 061;
V.~Bertin, E.~Nezri, and J.~Orloff, {\em JHEP} {\bf 02} (2003) 046;
R.~Arnowitt, B.~Dutta,  {{\tt  hep-ph/0204187}}; 
A.~Birkedal-Hansen,  hep-ph/0306144;
D.~G. Cerdeno and C.~Munoz, {{\tt hep-ph/0405057}}.

\bibitem{Belanger:micro13}
G.~B\'elanger, {\em et al.},
 {{\tt hep-ph/0405253}};
G.~B\'elanger,{\em et al.}, {\em Comput. Phys.
  Commun.} {\bf 149} (2002) 103--120.


\bibitem{Allanach:softsusy}
B.~C. Allanach, {\em Comput. Phys. Commun.} {\bf 143} (2002) 305--331,
  [\href{http://xxx.lanl.gov/abs/hep-ph/0104145}{{\tt hep-ph/0104145}}].

\bibitem{SLHA}
P.~Skands {\em et.~al.}, \href{http://xxx.lanl.gov/abs/hep-ph/0311123}{{\tt
  hep-ph/0311123}}.

\bibitem{Baer:reach}
H.~Baer, {\em et al.}, {\em JHEP} {\bf 06} (2003) 054;
H.~Baer {\em et.~al.}, {\em JHEP} {\bf 02} (2004)
  007.




\bibitem{stop}
C.~Boehm, A.~Djouadi, M.~Drees, {\em Phys, Rev.} {\bf D62} (2000) 035012;
Y.~Santoso, {\em Nucl. Phys. Proc. Suppl.} {\bf 124} (2003) 166--169.
 

  
\bibitem{Gomez:mb}
M.~E. Gomez, T.~Ibrahim, P.~Nath, and S.~Skadhauge,
  \href{http://xxx.lanl.gov/abs/hep-ph/0404025}{{\tt hep-ph/0404025}}.
  
  
\bibitem{codes}
A.~Djouadi {\em et al.}, hep-ph/0211331; W.~Porod, {\em Comput. Phys. Commun.} {\bf 153} (2003) 275; F.E.~Paige, {\em et al.}, hep-ph/0312045.

\bibitem{Allanach:houchesrge}
B.~C. Allanach, {\em et al.},
  \href{http://xxx.lanl.gov/abs/hep-ph/0402161}{{\tt hep-ph/0402161}}.


\bibitem{Atlas-tdr}
The Atlas Collaboration,  CERN-LHCC-99-14.


\bibitem{Baer:2004qq}
H.~Baer {\em et.~al.},
  \href{http://xxx.lanl.gov/abs/hep-ph/0405210}{{\tt hep-ph/0405210}}.

   
\end{thebibliography}
\end{document}